\documentclass[aps,prb,twocolumn,floatfix,showpacs]{revtex4}
\usepackage{amsmath}
\usepackage{graphicx}

\renewcommand{\r}{{\bf r}}
\renewcommand{\k}{{\bf k}}
\newcommand{\rhat}{{\bf \hat{r}}}
\newcommand{\khat}{{\bf \hat{k}}}

\begin{document}

\title{Optimal Fourier filtering of a function that is strictly confined 
       within a sphere}

\author{ Jos\'e M. Soler and Eduardo Anglada }
\affiliation{ Departamento de F\'{\i}sica de la Materia Condensada, 
              C-III, Universidad Aut\'{o}noma de Madrid,
              E-28049 Madrid, Spain }

\date{\today}

\begin{abstract}
   We present an alternative method to filter a distribution, that is 
strictly confined within a sphere of given radius $r_c$, so that its
Fourier transform is optimally confined within another sphere of
radius $k_c$.
   In electronic structure methods, it can be used to generate 
optimized pseudopotentials, pseudocore charge distributions, and
pseudo atomic orbital basis sets.
\end{abstract}



\maketitle


   In some computational problems we are interested in distributions
that are strictly confined within a sphere of given radius (i.\ e.\
defined to be strictly zero outside that sphere) and, simultaneously,
optimally confined within another sphere in reciprocal space, so that 
they can be well approximated by a finite number of Fourier components 
or, equivalently, by a finite number of grid points in real space.
   Within the field of electronic structure calculations, this 
typically occurs in the real-space application of pseudopotentials
~\cite{KingSmith-Payne-Lin1991,Briggs-Sullivan-Bernholc1996,
       Ono-Hirose1999,Wang2001,Tafipolsky-Schmid2006}.
   In the specific case of the SIESTA density functional method
\cite{Ordejon-Artacho-Soler1996,Soler2002}, this problem arises in the
evaluation, using a real-space grid, of matrix elements involving 
strictly localized basis orbitals~\cite{Sankey-Niklewski1989} and 
neutral-atom pseudopotentials.
   Those integrals produce an artificial rippling of the total energy, 
as a function of the atomic positions relative to the grid points
(the so-called ``egg box'' effect), which complicates considerably the 
relaxation of the geometry and the evaluation of phonon frequencies 
by finite differences.

   We have proposed recently a method to filter a distribution
simultaneously in real and reciprocal space~\cite{Anglada-Soler2006}.
   Such filter is optimal, in the sense of minimizing the norm of
the function outside two spheres of radius $r_c$ and $k_c$ in 
real and Fourier space, respectively.
   It works by projecting the distribution to be filtered on a basis
of functions that have the same shape in real and reciprocal space,
and that are thus optimally confined in both.
   However, because of the uncertainty principle, such basis functions,
and the resulting filtered distribution, cannot be {\em strictly} 
confined in any of the two spaces.
   Thus, if we insist in the strict confinement in real space,
and therefore we truncate the filtered pseudoatomic orbitals 
beyond $r_c$, they will have a discontinuity at $r_c$, and therefore 
an infinite kinetic energy.
   In practice, the smallness of the discontinuity, and the use of
integration grids with finite spacings, makes the problem more 
academic than real.
   But occasionally, when trying to converge the results to very
high precision, it is annoying to have such a potential problem.
   Another, independent problem in our previous procedure is that
the resulting functions, that were expanded in Legendre polynomials, 
do not obey exactly the correct behavior for $r \rightarrow 0$.
   In the present work, we propose an alternative method to filter
a distribution so that it is always strictly confined in real space,
while it is optimally confined in reciprocal space.


   Consider an initial function with a well defined angular momentum
and strictly confined within a sphere:
\begin{equation}
F_0(\r) = \left\{
  \begin{array}{ll} 
    F_0(r) Y_l^m(\rhat)  & \mbox{if $r \le r_c$} \\
    0                    & \mbox{otherwise,}
  \end{array}
\right.
\label{F0ofr}
\end{equation}
where $F_0(r)$ is continuous and $F_0(r_c)=0$.
   We are using the same symbol for $F_0(\r)$ and its radial
part $F_0(r)$, since it does not lead to any confusion.
   $Y_l^m(\rhat)$ is a real spherical harmonic.
   To require that $F(\r)$ [the filtered version of $F_0(\r)$] 
remains strictly zero for $r > r_c$, and continuous at $r_c$
(so that its kinetic energy is finite), we will expand it in terms
of spherical Bessel functions $j_l$ with a zero at $r_c$:
\begin{equation}
F(r) = \left\{
  \begin{array}{ll} 
    \sum_{n=1}^{M} c_n N_{ln} j_l(k_{ln} r)  & \mbox{if $r \le r_c$} \\
    0                             & \mbox{otherwise,}
  \end{array}
\right.
\label{Fofr}
\end{equation}
where $M$ is large enough to represent the function with the required 
accuracy, $k_{ln} r_c$ is the $n$th root of $j_l(x)$, and $N_{ln}$ are
normalization constants given by
\begin{equation}
N_{ln}^{-2} = \int_0^{r_c} r^2 dr ~j_l^2(k_{ln} r) = 
         \frac{r_c^3}{2} j_{l+1}^2(k_{ln} r_c).
\label{Nln}
\end{equation}

   The Fourier transform of $F(\r)$ is 
\begin{equation}
G(\k) \equiv \frac{i^l}{(2 \pi)^{3/2}} \int d^3\r ~e^{-i \k \r} F(\r) 
      = G(k) Y_l^m(\khat),
\label{Gofk}
\end{equation}
where we have introduced the factor $i^l$ to make $G(\k)$ real, and
\begin{equation}
G(k) = \sum_{n=1}^{N} G_n j_{ln}(k),
\label{Gofc}
\end{equation}
where $j_{ln}(k)$ is the Fourier transform of $N_{ln} j_l(k_{ln} r)$:
\begin{eqnarray}
j_{ln}(k) &\equiv& \frac{1}{(2 \pi)^{3/2}}  
              \int_0^{r_c} 4 \pi r^2 dr ~N_{ln} j_l(k_{ln} r) ~j_l(kr)
                                                          \nonumber \\
          &=& \left( \frac{r_c^3}{\pi} \right)^{1/2} \times \left\{
              \begin{array}{ll} 
                 j_{l+1}(k_{ln} r_c)          & \mbox{if $k=k_{ln}$} \\
                - \frac{2 k_{ln} r_c}{k^2 r_c^2 - k_{ln}^2 r_c^2} j_l(k r_c)
                                           & \mbox{otherwise.}
  \end{array}
\right.
\label{jlnofj}
\end{eqnarray}

\begin{figure}[htpb]
\includegraphics[width=0.9\columnwidth,clip]{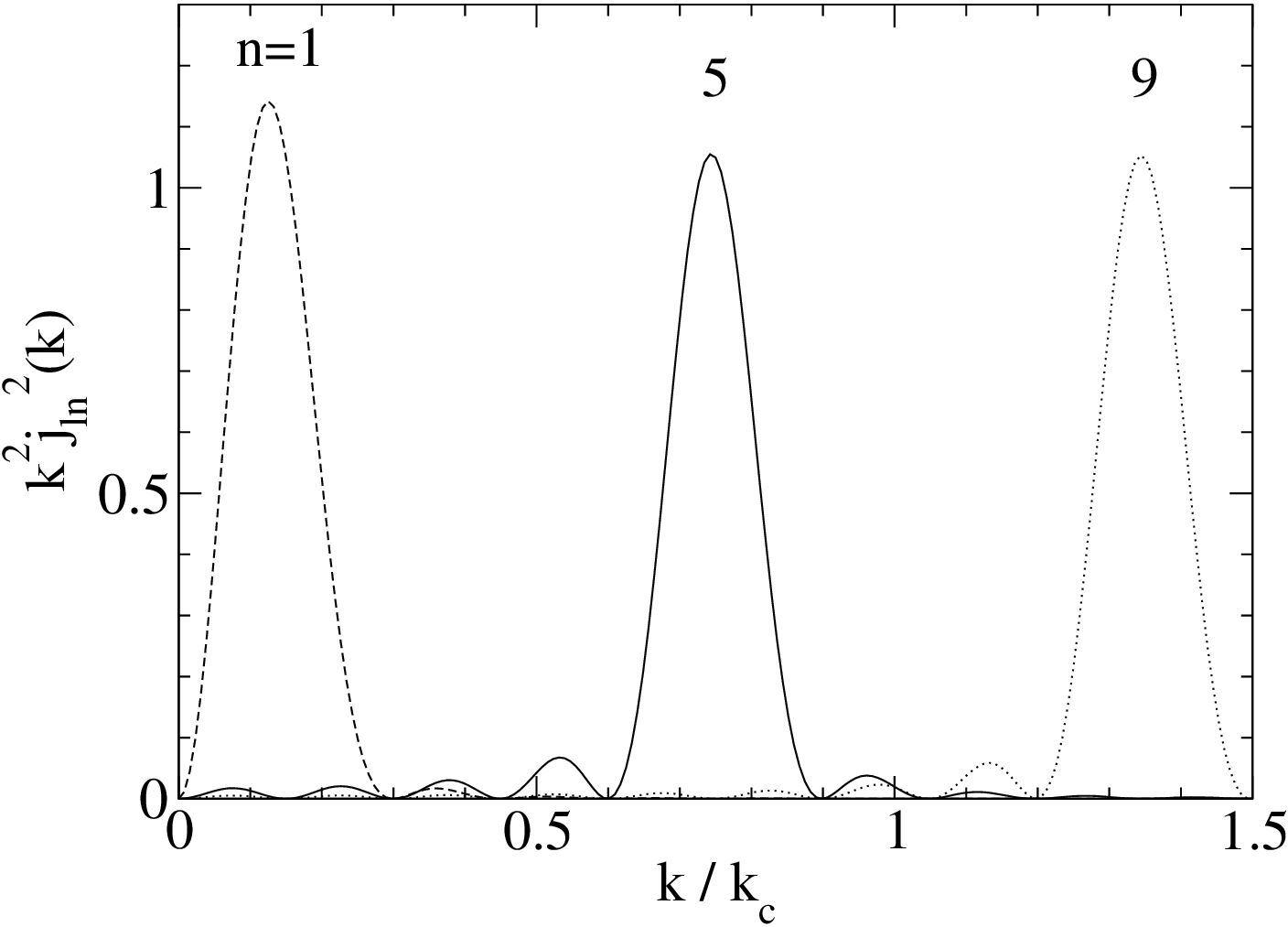}
\includegraphics[width=0.9\columnwidth,clip]{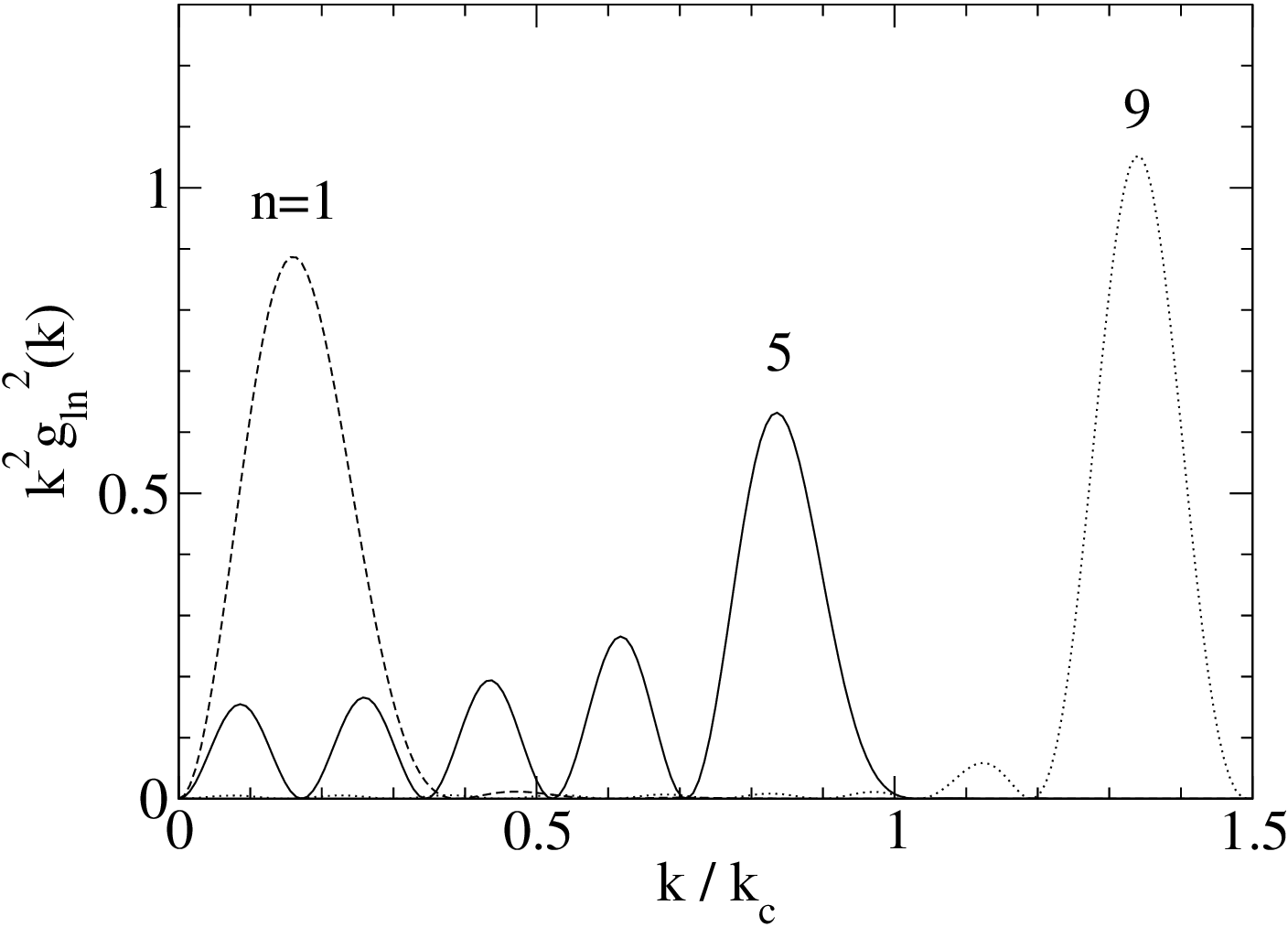}
\caption{
   Upper panel: squared Fourier transform, $k^2 j_{ln}^2(k)$, of 
some normalized spherical Bessel functions $j_l(k_{ln} r)$ strictly 
confined in $r \le r_c$.
   Lower panel: squared Fourier transform of selected solutions, 
$g_{ln}(k)$, to the problem of minimizing the kinetic energy in $k>k_c$,
Eq.~(\ref{Ekc}).
   $l=0, k_c r_c=25$.
}
\label{fig_jlnofk}
\end{figure}

   The basis functions $N_{ln} j_l(k_{ln} r)$ are the solutions to
Schr\"odinger's equation in a potential $V(r)=0$ for $r \le r_c$.
   According to the variational principle, they are the functions
that minimize the kinetic energy, among those strictly confined
within a sphere of radius $r_c$.
   Some of the Fourier transforms $j_{ln}(k)$ are shown in 
Fig.~\ref{fig_jlnofk}.
   They are delta-like functions in reciprocal space, broadened 
because of their confinement in real space, according to the 
uncertainty principle.

   A conventional and straightforward method to filter $F(\r)$ 
would be to project it on the basis $j_l(k_{ln} r)$, with $k_{ln} < k_c$,
i.\ e.\ by truncating the series in Eq.~(\ref{Fofr}).
   A better procedure is to use a basis of orthonormal functions 
that minimize not the total kinetic energy, but specifically the 
kinetic energy in the region $k>k_c$ that we want to filter out:
\begin{equation}
\int_{k_c}^\infty k^4 dk ~g_l^2(k) = \min.
\label{Ekc}
\end{equation}
   Expanding the solutions in the primitive basis,
\begin{equation}
g_l(k) = \sum_n c_n j_{ln}(k),
\label{gk}
\end{equation}
leads to the eigenvalue equation
\begin{equation}
   \sum_{m} H_{nm} c_m  = \epsilon c_n
\label{Hcmuc}
\end{equation}
where $\epsilon$ is a Lagrange multiplier to ensure normalization and
\begin{eqnarray}
H_{nm} &=& \int_{k_c}^\infty k^4 dk ~j_{ln}(k) ~j_{lm}(k) \nonumber \\
       &=& k_{ln}^2 \delta_{nm} - \int_0^{k_c} k^4 dk ~j_{ln}(k) ~j_{lm}(k).
\label{Hnm}
\end{eqnarray}
   The resulting eigenfunctions $g_{ln}(k)$ (that we will call 
``filterets'') are qualitatively very similar in real space to those 
in ref.[\onlinecite{Anglada-Soler2006}] and therefore they are not 
reproduced here again.
   Fig.~\ref{fig_jlnofk} shows them in reciprocal space for a very 
small value $k_c r_c =25$, used to emphasize the effects of an 
extreme confinement.
   When $k_{ln} << k_c$ or $k_{ln} >> k_c$, they are similar to 
the primitive functions $j_{ln}(k)$.
   For $k_{ln} \lesssim k_c$, however, they are considerably better 
confined within $k<k_c$.

\begin{figure}[htpb]
\includegraphics[width=0.9\columnwidth,clip]{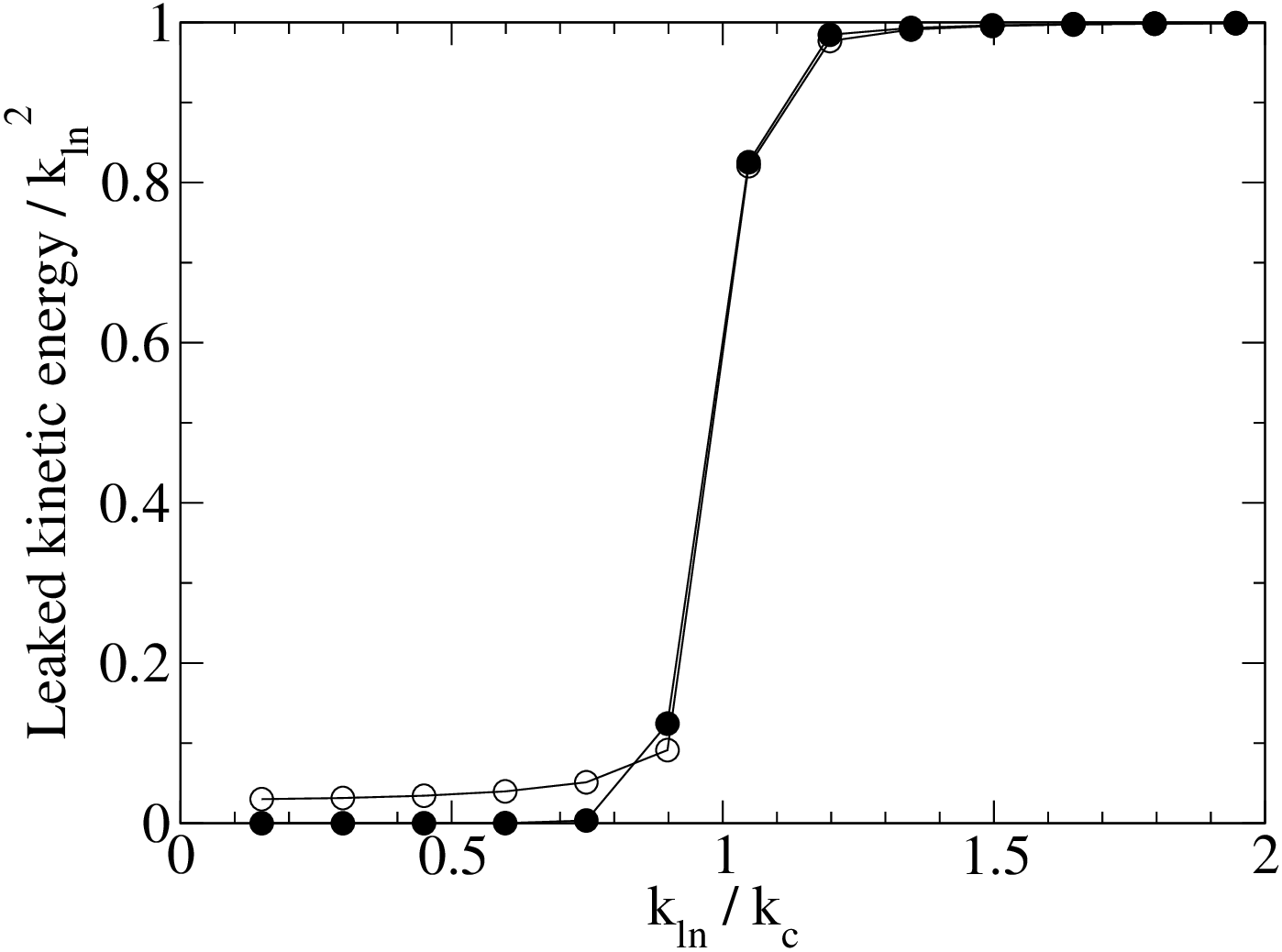}
\caption{
   Leaked kinetic energy, above the cutoff $k_c$, of the 
solutions $g_{ln}(k)$ to Eq.~(\ref{Ekc}) (eigenvalues 
$\epsilon_{ln}$, Eq.~(\ref{mun}), filled symbols) 
compared to the same integral $\int_{k_c}^{\infty} k^4 dk ~j_{ln}^2(k)$ 
for the confined spherical Bessel functions (empty symbols).
   $l=0, k_c r_c = 25$.
}
\label{fig_eigval}
\end{figure}

   The eigenvalues $\epsilon_{ln}$ of Eq.~(\ref{Hcmuc}) give the integral 
of the kinetic energy ``leaked'' outside $k_c$:
\begin{equation}
\epsilon_{ln} = \int_{k_c}^\infty k^4 dk ~g_{ln}^2(k).
\label{mun}
\end{equation}
   As expected, these eigenvalues 
are $\epsilon_{ln} \simeq 0$ for $k_{ln} < k_c$ and 
$\epsilon_{ln} \simeq k_{ln}^2$ for $k_{ln} > k_c$.
   They are compared in Fig.~\ref{fig_eigval} with the same integral
of the original functions and it can be seen that they are much 
smaller for $k_{ln} < k_c$.

   Since the functions $j_{ln}^2(k)$ minimize the {\em total}
kinetic energy, the decrease of kinetic energy in $k>k_c$ by
$g_{ln}^2(k)$  must be at the expense of a {\em larger}
increase in $k<k_c$, resulting in a net increase.
   To control this increase, we have found convenient to give
a small weight (say $w \sim 0.1$) to the kinetic energy in 
$k<k_c$.
   This can be done simply by multiplying the last integral in 
Eq.~\ref{Hnm} by a factor $(1-w)$.
   A very small value $w=10^{-6}$ was used in Fig.~\ref{fig_jlnofk},
just to break the degeneracy of the functions $g_{ln}(k)$ with 
$k_{ln}<<k_c$.
   Larger values yield functions somewhat intermediate between
both panels.

   The filtered function $F(\r)$ is then obtained by projecting
the original function $F_0(\r)$ over the subspace spanned by the
``filterets'' $g_{ln}^2(k)$ with a sufficiently low eigenvalue
(say $\epsilon_{ln}/k_{ln}^2 < 0.01$).
   We have checked that the resulting scheme produces pseudoatomic
orbitals, neutral-atom potentials~\cite{Soler2002}, and 
partial-core-correction densities~\cite{Louie-Froyen-Cohen1982:NLCC} 
that are free of the mentioned pathologies of the previous 
scheme~\cite{Anglada-Soler2006}, and that reduce the
``egg box'' effect in SIESTA at least as well.
   Overall, however, the convergence tests yield rather similar
results and therefore we do not repeat here the figures and tables
of reference~\onlinecite{Anglada-Soler2006}.

   Finally, a practical remark on the filtering procedure is appropriate.
   In grid-based methods~\cite{Beck2000}, in which the kinetic energy 
is calculated by finite differences, it is appropriate to use a 
filtering cutoff $k_c$ given by the maximum plane wave vector that
can be represented in the grid without 
aliasing~\cite{Briggs-Sullivan-Bernholc1996}.
   In SIESTA, however, the dominant kinetic energy is calculated by well
converged two-center integrals~\cite{Soler2002} that do not contribute
to the egg box effect.
   In this case, it is more convenient to fix $k_c$ by some independent
criterion, so that the orbitals (and the kinetic energy) do not depend 
on the integration grid used to calculate the exchange-correlation and 
pseudopotential interactions.
   Thus, we can fix an energy threshold $\epsilon_c$ such that
\begin{equation}
\epsilon_c = \int_{k_c}^\infty k^4 dk ~\phi^2(k),
\label{epsc}
\end{equation}
where $\phi$ are the original (unfiltered) atomic basis orbitals.
   This criterion will yield different (but balanced) reciprocal-space 
cutoffs $k_c$ for each orbital, in the same spirit that the 
``energy shift'' ~\cite{Soler2002} fixes their cutoffs $r_c$ in real 
space.
   The grid cutoff will then be fixed 
to $\sim 1.5-2$ times the maximum filter cutoff of all the orbitals 
(this factor coming from the fact that the plane wave cutoff for the 
density is larger than that for the wave functions).


   In conclusion, we have presented a simple but powerful method to 
generate a basis of orthonormal functions (``filterets''), with a given 
angular momentum, which are strictly confined within a cutoff radius in 
real space and optimally confined within another cutoff in Fourier space.
   We have described their use to filter a function that is 
strictly confined within a sphere.
   In addition, these orthonormal functions constitute themselves 
a general and systematically improvable basis for converged 
calculations using localized basis orbitals \cite{Haynes-Payne1997}.
   This possibility will be explored in future works.

\begin{acknowledgments}
   This work has been founded by grant FIS2006-12117 from the
Spanish Ministery of Science.
\end{acknowledgments}

\bibliographystyle{apsrev}
\bibliography{dft,siesta}

\end{document}